\documentclass[journal=jctcce,manuscript=article]{achemso}
\usepackage[version=3]{mhchem}

\usepackage{amssymb,amsmath,amsfonts,multicol,multirow,longtable,array}
\usepackage{lscape}
 
\usepackage{appendix}
\usepackage{soul} 
\usepackage[usenames]{color}
\usepackage{makecell}
\usepackage{enumerate}
\usepackage{xr}
\usepackage[T1]{fontenc} 
\usepackage{booktabs}
\usepackage{tikz}
\usepackage{threeparttable}
\usepackage{graphicx}
\usepackage[figuresright]{rotating}
\usepackage{amsfonts}
\usepackage{colortbl}
\usepackage{framed}
\usepackage{verbatim}
\usepackage{amsbsy}
\usepackage{bm}
\usepackage{bbm}
\usepackage[normalem]{ulem}
\usepackage{float}
\usepackage[linesnumbered,boxed]{algorithm2e}
\usepackage{algorithm2e}
\usepackage{diagbox}
\usepackage{tabularx} 
\usepackage{booktabs} 
\usepackage{physics}
\usepackage{subcaption}

\usepackage[colorlinks=true, urlcolor=blue]{hyperref}
\usepackage{enumitem}
\setlist[enumerate]{label=(\arabic*), itemsep=2pt, parsep=0pt, topsep=3pt, partopsep=0pt}
\usepackage{xcolor}

\usepackage{braket}

\newcounter{xformulation}

\newfloat{Algorithm}{htbp}{alg}
\floatname{Algorithm}{Algorithm}

\newcounter{exe}[figure]
\newcommand{\iexe}{\refstepcounter{exe}\the\value{exe}:}

\setkeys{acs}{maxauthors = 0} 

\author{Xiaoyu Zhang} 
\email{zhangxiaoyu@stu.pku.edu.cn} 
\affiliation{College of Chemistry and Molecular Engineering, Peking University, Beijing 100871, the People's Republic of China} 

\author{Taoni Bao}
\affiliation{School of Physics, Peking University, Beijing 100871, the People's Republic of China}
\alsoaffiliation{Center for Applied Physics and Technology, Peking University, Beijing 100871, the People's Republic of China}

\title{Operator Formalism for Noncollinear Functionals in Multicollinear Approach}

\begin{document}

\begin{abstract}
  Accurate modeling of spin-orbit coupling and noncollinear magnetism requires noncollinear density functionals within the two-component generalized Kohn-Sham (GKS) framework, yet constructing and implementing noncollinear functionals remains challenging. Recently, a well-defined methodology called the multicollinear approach was proposed to extend collinear functionals into noncollinear ones. While previous research focuses on its matrix representation, the present work derives its operator formalism. We implement these new equations in our noncollinear functional ensemble named NCXC, which is expected to facilitate compatibility with most DFT software packages. Since the multicollinear approach was proposed for solving nonphysical properties and mathematical singularities in noncollinear functionals, we validate its accuracy in practical periodic systems, including noncollinear magnetism in spin spirals, band structures in topological insulators, and band gaps in semiconducting inorganic materials.
\end{abstract}

\section{Introduction}
Density functional theory (DFT) \cite{KS-DFT} has garnered substantial attention due to its optimal balance between accuracy and efficiency. Considering that electron spin is crucial for the accurate description of systems, spin-DFT \cite{SpinDFT} has been developed to characterize spin-polarized systems, including ferromagnetic and antiferromagnetic materials. In three-dimensional real space, a magnetization vector $\bm{m}(\bm{r})$ is typically employed at each real space grid to represent the electron spin. When the magnetization vector is parallel or anti-parallel throughout the space, the system is referred to as collinear. In collinear systems, four components $n(\bm{r})$, $m_{x}(\bm{r})$, $m_{y}(\bm{r})$, $m_{z}(\bm{r})$ can be simplified to two components $n(\bm{r})$ and $m_{z}(\bm{r})$ or equivalently $\rho^{\uparrow}=\frac{n+m_{z}}{2}$ and $\rho^{\downarrow}=\frac{n-m_{z}}{2}$. Should $\bm{m}(\bm{r})$ not be parallel or anti-parallel to one another at individual points, the system is termed noncollinear. Similarly, the exchange and correlation functional in DFT can be classified into collinear and noncollinear functionals, represented by $E_{\mathrm{xc}}[n,m_z]$ and $E_{\mathrm{xc}}[n,\bm{m}]$, respectively. It is noteworthy that noncollinear functionals and noncollinear systems do not necessarily occur concurrently. For instance, spin-flip linear-response time-dependent density functional theory (TDDFT) of collinear systems requires noncollinear kernels \cite{RN69}.

There are two primary approaches for the development of noncollinear functionals. One particular approach involves the explicit formulation of a distinct noncollinear functional \cite{PhysRevLett.111.156401}, a process which exceeds the purview of the present study. An alternative approach lies in the extension of collinear functionals to noncollinear functionals, which is both more feasible and simpler for standard DFT calculations. In 1988, Kübler \emph{et al.} \cite{Kubler_1988} were the pioneers in extending the local spin density approximation (LSDA) \cite{LSDA} by substituting $\bm{m}$ for $|\bm{m}|$. This methodology is identified as the locally collinear approach herein, as it entails the projection of $\bm{m}$ onto its inherent direction. The locally collinear approach is theoretically designed for noncollinear LSDA \cite{Kubler_1988} and it has been computationally elaborated to noncollinear generalized gradient approximation (GGA) and noncollinear meta-GGA (mGGA) \cite{PhysRevB.75.125119}. Nonetheless, the locally collinear approach encounters several theoretical and computational challenges beyond those associated with LSDA. For exchange-correlation (xc) potential, serious numerical problems have been reported \cite{10.1063/5.0051447,10.1063/5.0005094}. Eq. (68) in Ref. \citenum{10.1063/5.0005094} and Eqs. (47)-(50) in Ref. \citenum{10.1063/5.0051447} drop some unstable terms to circumvent numerical problems. Moreover, the lack of non-vanishing local torque is also reported \cite{PhysRevLett.98.196405}. For xc kernel, theoretical and numerical problems are collectively reviewed, and no general truncation has been found yet in Ref. \citenum{C8CS00175H} Section C. In 2012, Giovanni Scalmani and Michael J. Frisch \cite{doi:10.1021/ct300441z} introduced a generalization applicable to both GGA \cite{PhysRevB.46.6671,PhysRevA.38.3098} and meta-GGA \cite{10.1063/1.454274,doi:10.1139/v89-073,NEUMANN199716} by constructing auxiliary inputs, which are also found to give a numerically unstable potential \cite{10.1063/5.0051447} beyond LSDA. In the context of TDDFT, artificial truncation is necessary and lacking global spin rotation invariance after truncation is reported. \cite{doi:10.1021/acs.jctc.7b00104} Lacking a general solution for noncollinear potential and kernel functionals necessitates a proposal of a unified and well-defined theory which doesn't need any artificial truncation. More recently, in 2023, Zhichen Pu \emph{et al.} presented a generalization termed the multicollinear approach \cite{PhysRevResearch.5.013036}, which is applicable to any type of exchange and correlation functional. The multicollinear approach proposes to project $\bm{m}$ onto different sampling directions rather than its inherent direction. The multicollinear approach basically provides a well-defined noncollinear functional without any additional numerical truncation, which is reported to satisfy the following four properties \cite{PhysRevResearch.5.013036}:
\begin{enumerate}
    \item Meeting the correct collinear limit;
    \item Invariant to global spin rotations while sensitive to local spin rotations;
    \item Having well-defined functional derivatives;
    \item Providing global zero torque but non-vanishing local torque.
\end{enumerate}
Here, we also mention that for orbital-dependent meta-GGA, a different generalization scheme preserving the U(1) $\times$ SU(2) symmetry has been developed, leading to a spin-current density functional \cite{PhysRevB.96.035141,PhysRevLett.133.136401}, which is outside the scope of this paper. The multicollinear approach eliminates the requirement for numerical screening and has not demonstrated instability in its terms; consequently, this method appears promising for prospective extension to incorporate higher derivatives than xc kernels, such as the two-component TDDFT analytical gradients \cite{doi:10.1021/acs.jctc.5c00115}.

In prior studies \cite{PhysRevResearch.5.013036,RN69,doi:10.1021/acs.jctc.5c00115}, the multicollinear approach was predominantly explored through matrix representations, adhering to an algebraic methodology, wherein the explicit form of the exchange and correlation potential and kernel operator was not articulated. Furthermore, in the case of noncollinear semi-local functionals, the incorporation of the derivatives of basis functions into these matrix representations obstructs the direct derivation of the multiplicative noncollinear xc potential and kernel operator, even when considering noncollinear GGA. This scenario diverges considerably from the context of collinear functionals, stemming from the inherent complexity of the multicollinear approach. Consequently, deriving the operator formalism and making it reduce to the multiplicative operator in semi-local approximation for noncollinear functionals within the multicollinear framework is a nontrivial task. The establishment of an operator formalism offers several advantages:
\begin{enumerate}
    \item It achieves a theoretical framework consistent with existing generalization methods, such as the locally collinear approach and the Scalmani-Frisch method, and also reaches a consistency with existing collinear functional operators;
    \item It enables the implementation of a noncollinear functional ensemble that collectively provides exchange and correlation potentials and, in principle, higher functional derivatives in the future;
    \item It facilitates more adaptable future developments in both orbital-based and grid-based computational packages.
\end{enumerate}
Motivated by this, this study achieves three advancements:
\begin{enumerate}
    \item From a theoretical standpoint, we derive the operator formalism for noncollinear functionals in the multicollinear approach;
    \item On the computational front, our method is implemented into a noncollinear functional ensemble, designated as NCXC;
    \item In terms of application, we confirm the accuracy of this approach for practical periodic systems, noting that the multicollinear approach has not previously been validated within these systems.
\end{enumerate}

\section{Theory and Implementation}

\subsection{Two-Component GKS Framework}

We denote two-component basis and orbitals as follows:

\begin{enumerate}
    \item General two-component basis set: Allowing the mixing of spin-up and spin-down basis sets, represented by Greek letters $\Gamma, \Lambda, \Theta$;
    \item General two-component molecular orbitals or $\bm{k}$-dependent crystal orbitals: Labeled with uppercase English letters $P, Q, R$, usually obtained from generalized Kohn-Sham (GKS) calculations;
    \item Occupied general two-component molecular orbitals or $\bm{k}$-dependent crystal orbitals: Labeled with $I, J, K$;
    \item  In this study, if under periodic boundary condition, all bases, orbitals and orbital coefficients exhibit dependence on the wave vector $\bm{k}$; however, for the sake of simplicity, we shall omit the explicit labeling of $\bm{k}$.
\end{enumerate}

It is important to acknowledge that we shall adhere to the Einstein summation convention for the sake of simplicity.
Within the two-component GKS formalism, the orbitals are expressed in terms of two-component bases,
\begin{equation}
    P = C_{\Lambda P} \Lambda
\end{equation}
where $C_{\Lambda P}$ are the expansion coefficients of the orbital $P$ in the basis $\Lambda$.

The first-order reduced density matrix is defined as follows:
\begin{equation}
   D_{\Lambda \Gamma} =  C_{\Lambda I} C_{\Gamma I}^{*}
\end{equation}

The matrix representation of the Hamiltonian for two-component generalized Kohn-Sham theory gives,
\begin{equation}
\begin{split}
    H_{\Gamma \Lambda}& = h_{\Gamma \Lambda} + (\Gamma \Lambda | \Pi \Theta)D_{\Theta \Pi} -c_{\mathrm{HF}}
    (\Gamma \Theta | \Pi \Lambda)D_{\Theta \Pi} \\
   &\quad + \int \frac{\delta E_{\mathrm{xc}}}{\delta n} \Gamma^{\dagger} \Lambda \,\mathrm{d}\bm{r} + \int\Gamma^{\dagger} \bm{\sigma}\cdot \frac{\delta E_{\mathrm{xc}}}{\delta \bm{m}} \Lambda \,\mathrm{d}\bm{r}
    \end{split}
\end{equation}
In this study, the two-component GKS is employed to extend the KS equation into a two-component framework \cite{PhysRevB.75.125119} as opposed to determining a non-multiplicative potential operator for orbital-dependent mGGA \cite{PhysRevB.53.3764}. Within the non-relativistic limit (nrl), the one-electron operator $\hat{h}$ is expressed as,
\begin{equation}
    \hat{h}^{\mathrm{nrl}} = \frac{\bm{p}^2}{2} \hat{I} +V_{\mathrm{ext}}
\end{equation}
To demonstrate that the two-component GKS constitutes a suitable framework for incorporating full relativity, we expound upon certain aspects of special relativity. Analytical emphasis is commonly placed on the transformation of the free-particle Dirac equation, as opposed to the Dirac-Coulomb-Breit Hamiltonian. \cite{Liu10072010} In other words, special relativity is usually applied to the one-electron operator $\hat{h}$. Starting from the four-component free-particle Dirac equation, a series of two-component full relativistic one-electron terms are derived, such as zeroth order regular approximation (ZORA) \cite{https://doi.org/10.1002/(SICI)1097-461X(1996)57:3<281::AID-QUA2>3.0.CO;2-U} and exact two-component (X2C) transformation \cite{Liu10072010}. For example, the explicit form of the ZORA operator is expressed as:
\begin{equation}
    \hat{h}^{\mathrm{ZORA}} = \bm{\sigma}\cdot \bm{p} \frac{c^2}{2c^2-V} \bm{\sigma}\cdot \bm{p} +V_{\mathrm{ext}}
\end{equation}
ZORA is originally derived from the single-particle Dirac equation, so the two potentials in the expression are originally the external potentials \cite{https://doi.org/10.1002/(SICI)1097-461X(1996)57:3<281::AID-QUA2>3.0.CO;2-U}. For KS-DFT, Van W\"ullen proposed that the $V$ in $2c^2-V$ should be the KS actual potential. However, it obtains fairly large errors from such a procedure unless special precautions are taken. Thus, Van W\"ullen thinks that it is reasonable to replace the $V$ by an approximated model potential \cite{10.1063/1.476576}. Different from ZORA, X2C is an algebraic method and its operator formalism hasn't been found, but it can still be processed in the matrix representation of two-component GKS \cite{10.1063/1.4758987,doi:10.1021/acs.jctc.3c00347}.

In addition to transforming the Dirac equation from a four-component to a two-component form, an alternative, more approximate approach involves the utilization of a pseudopotential engineered to align with full special relativity. The Kleinman-Bylander formalism \cite{kleinman1982efficacious}, a widely accepted approach, is employed within the density functional theory package utilized in this study, ABACUS. This method effectively decomposes the external potential into two distinct components: the scalar-multiplicative local component $\hat{V}^{\mathrm{L}}(\bm{r})$ and the non-diagonal nonlocal component $\hat{V}^{\mathrm{NL}}(\bm{r},\bm{r}^{\prime})$, which incorporates spin-orbit coupling. Both components are compatible with the two-component GKS framework. Specifically, the nonlocal component is described as \cite{PhysRevB.21.2630,hemstreet1993first,theurich2001self}:
\begin{equation}
    \hat{V}^{\mathrm{NL}} = E^I_{ l,j,\tau} (\beta^I_{ l,j,\tau}\mathrm{Y}^{j,m_j}_{l,1/2})  (\beta^I_{ l,j,\tau}\mathrm{Y}^{j,m_j}_{l,1/2})^{\dagger}
\end{equation}
where $I$ is the atom index, $(l,j,m_j)$ is a set of quantum numbers, $\beta_\tau$ is the $\tau$-th Kleinman-Bylander projector, $\mathrm{Y}$ is the two-component spherical harmonic spinor, $E$ is the channel energy.

\subsection{Noncollinear Potential Operator}
Based on the above subsection, the noncollinear exchange-correlation potential operator can be expressed as:
\begin{equation}
    \hat{V}_{\mathrm{xc}} = \frac{\delta E_{\mathrm{xc}}[n,\bm{m}]}{\delta n} \hat{I} + \frac{\delta E_{\mathrm{xc}}[n,\bm{m}]}{\delta \bm{m}}\cdot \bm{\sigma}
\end{equation}
However, noncollinear functionals $E_{\mathrm{xc}}[n,\bm{m}]$ haven't been widely developed yet, so we use the multicollinear approach to deduce noncollinear functionals from collinear functionals $E_{\mathrm{xc}}[\rho^{\uparrow},\rho^{\downarrow}]$. We obtain that:
\begin{equation}
    \hat{V}_{\mathrm{xc}} = \frac{1}{4 \pi} \int_0^{2\pi} \int_0^\pi \left[\left.\frac{\delta E_{\mathrm{xc}}^{\mathrm{eff}}[n,s]}{\delta n}\right|_{s=\bm{m}\cdot \bm{e}}\hat{I}+
    \left.\frac{\delta E_{\mathrm{xc}}^{\mathrm{eff}}[n,s]}{\delta s}\right|_{s=\bm{m}\cdot \bm{e}}(\bm{\sigma} \cdot \bm{e})
    \right] \sin \theta \,\mathrm{d} \theta \,\mathrm{d}\phi
\end{equation}
where
\begin{equation}
    \bm{e}(\theta,\phi) = (\sin \theta \cos \phi, \sin \theta \sin \phi, \cos \theta)
\end{equation}
and
\begin{subequations}
    \begin{equation}
    \begin{split}
        \frac{\delta E_{\mathrm{xc}}^{\mathrm{eff}}[n,s]}{\delta n(\bm{r}_1)} &= \frac{1}{2}\left(\frac{\delta E_{\mathrm{xc}}[\rho^{\uparrow},\rho^{\downarrow}]}{\delta \rho^{\uparrow}(\bm{r}_1)}+\frac{\delta E_{\mathrm{xc}}[\rho^{\uparrow},\rho^{\downarrow}]}{\delta \rho^{\downarrow}(\bm{r}_1)}\right)
        +\frac{1}{4} \int \mathrm{d}\bm{r}_2\left(\rho^{\uparrow}(\bm{r}_2)-\rho^{\downarrow}(\bm{r}_2)\right)\\
        &\quad\left(\frac{\delta^2 E_{\mathrm{xc}}[\rho^{\uparrow},\rho^{\downarrow}]}{\delta \rho^{\uparrow}(\bm{r}_1)\delta \rho^{\uparrow}(\bm{r}_2)}-\frac{\delta^2 E_{\mathrm{xc}}[\rho^{\uparrow},\rho^{\downarrow}]}{\delta \rho^{\uparrow}(\bm{r}_1)\delta \rho^{\downarrow}(\bm{r}_2)}+\frac{\delta^2 E_{\mathrm{xc}}[\rho^{\uparrow},\rho^{\downarrow}]}{\delta \rho^{\downarrow}(\bm{r}_1)\delta \rho^{\uparrow}(\bm{r}_2)}-\frac{\delta^2 E_{\mathrm{xc}}[\rho^{\uparrow},\rho^{\downarrow}]}{\delta \rho^{\downarrow}(\bm{r}_1)\delta \rho^{\downarrow}(\bm{r}_2)}\right)
    \end{split}
    \end{equation}
    \begin{equation}
    \begin{split}
        \frac{\delta E_{\mathrm{xc}}^{\mathrm{eff}}[n,s]}{\delta s(\bm{r}_1)} &= \frac{\delta E_{\mathrm{xc}}[\rho^{\uparrow},\rho^{\downarrow}]}{\delta \rho^{\uparrow}(\bm{r}_1)}-\frac{\delta E_{\mathrm{xc}}[\rho^{\uparrow},\rho^{\downarrow}]}{\delta \rho^{\downarrow}(\bm{r}_1)}
        +\frac{1}{4} \int \mathrm{d}\bm{r}_2\left(\rho^{\uparrow}(\bm{r}_2)-\rho^{\downarrow}(\bm{r}_2)\right)\\
        &\quad\left(\frac{\delta^2 E_{\mathrm{xc}}[\rho^{\uparrow},\rho^{\downarrow}]}{\delta \rho^{\uparrow}(\bm{r}_1)\delta \rho^{\uparrow}(\bm{r}_2)}-\frac{\delta^2 E_{\mathrm{xc}}[\rho^{\uparrow},\rho^{\downarrow}]}{\delta \rho^{\uparrow}(\bm{r}_1)\delta \rho^{\downarrow}(\bm{r}_2)}-\frac{\delta^2 E_{\mathrm{xc}}[\rho^{\uparrow},\rho^{\downarrow}]}{\delta \rho^{\downarrow}(\bm{r}_1)\delta \rho^{\uparrow}(\bm{r}_2)}+\frac{\delta^2 E_{\mathrm{xc}}[\rho^{\uparrow},\rho^{\downarrow}]}{\delta \rho^{\downarrow}(\bm{r}_1)\delta \rho^{\downarrow}(\bm{r}_2)}\right)
    \end{split}
    \end{equation}
\end{subequations}
From a conceptual perspective, the locally collinear approach and the Scalmani-Frisch formulation exhibit numerical instability due to the presence of $|\bm{m}|$ in their exchange-correlation potential and kernel denominator. The equations presented herein address these numerical issues by substituting $|\bm{m}|$ with $\bm{m}\cdot \bm{e}$. A graphical elucidation involves projecting $\bm{m}$ onto each sampling direction and computing each collinear contribution to noncollinear functionals. The current equations, in principle, offer noncollinear potential operators applicable to any density functional, though the process might be implicit for orbital-dependent functionals or more complicated functionals. However, for computational implementation and integration efficiency, it becomes essential to derive specific equations, especially for the kernel $\dfrac{\delta^2 E_{\mathrm{xc}}[\rho^{\uparrow},\rho^{\downarrow}]}{\delta \rho^{\sigma}(\bm{r}_1)\delta \rho^{\sigma^{\prime}}(\bm{r}_2)}$. This necessitates exploiting particular features of each functional type to facilitate further derivations. Apparently, these equations are directly applicable to nonlocal functionals with an integrand dependent on $\bm{r}_1$ and $\bm{r}_2$, such as the weighted spin-density approximation (WSDA) \cite{PhysRevA.48.4197}. Recently, Tai Wang \emph{et al.} developed a similar set of equations only adapted for the nonlocal functional WSDA \cite{10.1063/5.0260762}. In this work, our existing equations are directly applicable to WSDA. Additionally, we discuss more commonly employed functionals including LDA, GGA, and mGGA. For purely local or semilocal functionals, like LDA, GGA, and Laplacian-dependent mGGA, we express their kernels as:
\begin{subequations}
    \begin{equation}
        \frac{\delta^2 E^{\mathrm{LDA}}_{\mathrm{xc}}[\rho^{\uparrow},\rho^{\downarrow}]}{\delta \rho^{\sigma}(\bm{r}_1)\delta \rho^{\sigma^{\prime}}(\bm{r}_2)} = \frac{\delta^2 E^{\mathrm{LDA}}_{\mathrm{xc}}[\rho^{\uparrow},\rho^{\downarrow}]}{\delta \rho^{\sigma}(\bm{r}_1)\delta \rho^{\sigma^{\prime}}(\bm{r}_1)} \delta(\bm{r}_1-\bm{r}_2)
    \end{equation}
    \begin{equation}
    \begin{split}
        \frac{\delta^2 E^{\mathrm{GGA}}_{\mathrm{xc}}[\rho^{\uparrow},\rho^{\downarrow}]}{\delta \rho^{\sigma}(\bm{r}_1)\delta \rho^{\sigma^{\prime}}(\bm{r}_2)}& = \frac{\partial}{\partial \rho^{\sigma^{\prime}}(\bm{r}_2)}\frac{\delta E^{\mathrm{GGA}}_{\mathrm{xc}}[\rho^{\uparrow},\rho^{\downarrow}]}{\delta \rho^{\sigma}(\bm{r}_1)}\delta(\bm{r}_1-\bm{r}_2)\\
        &\quad +\left[\frac{\partial}{\partial \nabla \rho^{\sigma^{\prime}}(\bm{r}_2)}\frac{\delta E^{\mathrm{GGA}}_{\mathrm{xc}}[\rho^{\uparrow},\rho^{\downarrow}]}{\delta \rho^{\sigma}(\bm{r}_1)}\right]\cdot \nabla \delta(\bm{r}_1-\bm{r}_2)
         \end{split}
    \end{equation}
    \begin{equation}
    \begin{split}
        \frac{\delta^2 E^{\mathrm{mGGA}}_{\mathrm{xc}}[\rho^{\uparrow},\rho^{\downarrow}]}{\delta \rho^{\sigma}(\bm{r}_1)\delta \rho^{\sigma^{\prime}}(\bm{r}_2)} &= \frac{\partial}{\partial \rho^{\sigma^{\prime}}(\bm{r}_2)}\frac{\delta E^{\mathrm{mGGA}}_{\mathrm{xc}}[\rho^{\uparrow},\rho^{\downarrow}]}{\delta \rho^{\sigma}(\bm{r}_1)}\delta(\bm{r}_1-\bm{r}_2)\\
        &\quad +\left[\frac{\partial}{\partial \nabla \rho^{\sigma^{\prime}}(\bm{r}_2)}\frac{\delta E^{\mathrm{mGGA}}_{\mathrm{xc}}[\rho^{\uparrow},\rho^{\downarrow}]}{\delta \rho^{\sigma}(\bm{r}_1)}\right]\cdot \nabla \delta(\bm{r}_1-\bm{r}_2)\\
        &\quad +\left[\frac{\partial}{\partial \nabla^2 \rho^{\sigma^{\prime}}(\bm{r}_2)}\frac{\delta E^{\mathrm{mGGA}}_{\mathrm{xc}}[\rho^{\uparrow},\rho^{\downarrow}]}{\delta \rho^{\sigma}(\bm{r}_1)}\right]\cdot \nabla^2 \delta(\bm{r}_1-\bm{r}_2)
        \end{split}
    \end{equation}
\end{subequations}
Semi-local approximation intuitively should have multiplicative xc operators, and we need to reach this goal in our operator formalism here. Direct substitution by the above expressions doesn't reach a multiplicative operator in semi-local approximation. Then, we need to utilize integration by parts in conjunction with the chain rule, and the multiplicative noncollinear potential can be articulated through the use of density tensors and the partial differentiation of the exchange-correlation energy integrand. For orbital-dependent mGGA, the implicit derivation of its multiplicative noncollinear potential is achievable through the application of the noncollinear optimal effective potential (OEP) \cite{PhysRevB.98.035140}. In order to make our expressions more accurate, it is imperative to elucidate the concept of the multiplicative operator. The term `multiplicative' is frequently employed within a one-component framework; for instance, a scalar potential is regarded as multiplicative, whereas $\nabla$ is not classified as such. Given that our study adopts a two-component framework, it becomes essential to broaden the definition to encompass matrix multiplication. Appendix A offers a mathematical delineation of the multiplicative property of operators in the context of a two-component framework.

\subsection{Extension Beyond Noncollinear Potential}
\subsubsection{XC Magnetic Field and XC Torque}
An anticipated characteristic of the precise xc magnetic field in noncollinear DFT, represented as $\bm{B}_{\mathrm{xc}}=\dfrac{\delta E_{\mathrm{xc}}[n,\bm{m}]}{\delta \bm{m}}$, is its capability to produce a local nonzero xc torque $-\bm{m}\times\bm{B}_{\mathrm{xc}}$. This occurs because the Kohn-Sham current diverges from the actual many-body current. On an overall scale, this xc torque must sum to zero, a requirement referred to as the zero-torque theorem \cite{Capelle2001SpinCA}. The xc torque plays an indispensable role in combining the Landau-Lifshitz-Gilbert (LLG) equation with DFT \cite{HUANG2023171098}, which has been widely used for simulating spin dynamics in spintronics \cite{2022AdM3400327Y}. LLG is a semi-classical method that follows a macroscopic equation and has to refine some empirical parameters. Also, the xc torque is found to be one important term in the time evolution equation of $\bm{m}(\bm{r},t)$ \cite{doi:10.1021/acs.jctc.4c01218}, 
\begin{equation}
    \frac{1}{2}\left(\frac{\mathrm{d}\bm{m}(\bm{r},t)}{\mathrm{d}t}+\nabla \cdot \bm{J}_{\mathrm{KS}}(\bm{r},t)\right) = -\bm{m}(\bm{r},t)\times\bm{B}_{\mathrm{xc}}(\bm{r},t)
\end{equation}
where the divergence of Kohn-Sham spin current is given by
\begin{equation}
    \nabla \cdot \bm{J}_{\mathrm{KS}} = \frac{1}{2\mathrm{i}} \sum_{i\in \mathrm{occ}} [\psi_i^{\dagger}(\bm{r},t) \hat{\sigma} \nabla^2 \psi_i(\bm{r},t)] + \mathrm{c.c.}
\end{equation}
c.c. is short for complex conjugate.

For LSDA, the exchange-correlation torque is inherently expected to be zero; however, this is not the case in more advanced approximations beyond LSDA. Various approaches have been pursued to achieve a non-vanishing local xc torque, such as Scalmani and Frisch's adaptations to the locally collinear approach \cite{PhysRevB.87.035117}, and the interplay of exact exchange with an optimized effective potential by Sharma \emph{et al.}\cite{PhysRevLett.98.196405}. Ref. \citenum{PhysRevB.87.035117} is a direct application of Scalmani-Frisch formulation so it suffers from the drawbacks we mentioned before. Ref. \citenum{PhysRevLett.98.196405} needs OEP and Hartree-Fock exchange, which is not applicable to pure functionals and has heavy computational costs. In the context beyond LSDA, it is anticipated that the xc torque within the multicollinear approach will exhibit a non-zero value. Thanks to our operator formalism, we can easily evaluate the xc magnetic field and torque. To elucidate this phenomenon, we present a toy GGA functional. By positing the xc energy integrand as $f=\sum_{\alpha,\beta=x,y,z}\nabla_{\alpha}m_{\beta}\nabla_{\alpha}m_{\beta}$, the formulation leads to $\bm{B}_{\mathrm{xc}}=2\nabla^2 \bm{m}$. Furthermore, under the assumption $\bm{m}=(x^2,2,3)$, the resulting xc local torque is $(0,6,-4)$, confirming its non-zero nature.

\subsubsection{Noncollinear Kernel for Spin-Flip TDDFT}

Upon its initial proposal by Yihan Shao \emph{et al.}\cite{10.1063/1.1545679}, spin-flip TDDFT utilized collinear exchange-correlation kernels. The coupling between molecular orbitals with different spins was facilitated by the Hartree-Fock (HF) exchange component within hybrid functionals. However, significant limitations are associated with spin-flip TDDFT employing collinear kernels. Firstly, in order to ensure accurate inter-spin channel coupling, there is typically a requisite reliance on hybrid functionals incorporating a substantial proportion of HF exchange, thereby limiting the appropriate use of pure functionals. Secondly, it is subject to the self-splitting problem, wherein the energies of the reference state $\ket{T(S_z=1)}$ and the spin-flip state $\ket{T(S_z=0)}$ are artificially disparate, resulting in ambiguity and inconsistency. Thirdly, it manifests spin contamination stemming from three sources: the inadequate treatment of spin response by collinear functionals, the incomplete representation of spin space, and the inherent contamination originating from the reference triplet state.

Conversely, spin-flip TDDFT utilizing noncollinear kernels, as initially posited by Wang and Ziegler \cite{10.1063/1.1844299}, constitutes a more rigorous alternative. Within this theoretical paradigm, pure functionals are capable of directly coupling the two spin channels \cite{10.1063/1.4714499}, thereby rendering the HF exchange an optional rather than an indispensable component. The spin-flip kernel introduced by Wang and Ziegler was originally founded on a noncollinear LSDA functional and is, in fact, the second derivative of noncollinear functionals from the locally collinear approach. When this kernel is extended to functionals beyond the LSDA, it encounters analogous numerical instabilities to those observed in the locally collinear approach, attributed to divisions by zero, thus constraining its broad applicability \cite{10.1063/1.3676736}. Noncollinear kernels in the matrix representation derived from the multicollinear approach have demonstrated numerical stability and are capable of resolving the self-splitting issue beyond the LDA with precision \cite{doi:10.1021/acs.jctc.5c00714}.

Commencing from our established noncollinear potential, the resultant derivation yields the noncollinear kernel applicable to spin-flip TDDFT:
\begin{equation}
\begin{split}
    \hat{K}_{\mathrm{xc}} &= \frac{1}{4 \pi} \int_0^{2\pi} \int_0^\pi \bigg[\frac{\delta^2 E^{\mathrm{eff}}[n,s]}{\delta n(\bm{r}_1) \delta n(\bm{r}_2)} \hat{I}+ \frac{\delta^2 E^{\mathrm{eff}}[n,s]}{\delta n(\bm{r}_1) \delta s(\bm{r}_2)} \bm{e} \cdot \bm{\sigma}(\bm{r}_2) \\
    &\quad +\bm{e} \cdot \bm{\sigma}(\bm{r}_1) \frac{\delta^2 E^{\mathrm{eff}}[n,s]}{\delta s(\bm{r}_1) \delta n(\bm{r}_2)}  +[\bm{e} \cdot \bm{\sigma}(\bm{r}_1)]\frac{\delta^2 E^{\mathrm{eff}}[n,s]}{\delta s(\bm{r}_1) \delta s(\bm{r}_2)}  [\bm{e} \cdot \bm{\sigma}(\bm{r}_2)]
    \left.\bigg]\right|_{s=\bm{m}\cdot\bm{e}} \sin \theta \,\mathrm{d} \theta \,\mathrm{d} \phi
    \end{split}
\end{equation}
Analogously, this equation is applicable to all types of density functionals; however, the mathematical properties inherent in each collinear functional are necessary to proceed further. 

\section{Numerical Tests and Computational Details}
\subsection{Computational Details}
Our noncollinear functional ensemble is released on GitHub \cite{ncxc}. In NCXC, the input concerning $\nabla^k  n$ and $\nabla^k  \bm{m}$ and output concerning the zeroth and first-order derivatives are described as follows:
\begin{enumerate}
    \item For noncollinear LSDA, the input is $n$ and $\bm{m}$, while the output is xc energy density per electron and $2\times2$ xc potential matrix;
    \item For noncollinear GGA, the input is order-0, order-1, order-2 $n$ and $\bm{m}$ gradients, while the output is  xc energy density per electron and $2\times2$ xc potential matrix;
    \item For noncollinear laplacian-dependent mGGA, the input is order-0, order-1, order-2, order-3 $n$ and $\bm{m}$ gradients, while the output is  xc energy density per electron and $2\times2$ xc potential matrix.
\end{enumerate}
Exchange and correlation energy can be calculated by integrating the multiplication of charge density and xc energy density per electron over the whole real space. The xc part Hamiltonian can be calculated by integrating the multiplication of two corresponding bases and the pure xc potential matrix. NCXC has nothing to do with Hartree-Fock exchange in hybrid functionals \cite{10.1063/1.464913} and dispersion correction \cite{10.1063/1.4993215,10.1063/1.5090222} but simply does addition. In NCXC, two distinct methodologies for additional spin sampling have been incorporated: Lebedev sampling and Fibonacci sampling. By default, NCXC utilizes Lebedev sampling, with the default configuration set to 1454 Lebedev grids. It is important to emphasize that the sampling in this context refers specifically to additional spin sampling, rather than atomic real space sampling for integration.

We establish a connection between NCXC and a numerical atomic orbital (NAO)-based density functional theory open-source software package, ABACUS (Atomic-orbital Based Ab-initio Computation at USTC) \cite{LI2016503,https://doi.org/10.1002/wcms.1687}. All noncollinear computations in this study are conducted by employing GKS calculations within ABACUS, except the spin spiral calculation, which is carried out within FLEUR \cite{fleurWeb}. Settings regarding pseudopotentials, crystal structure, NAO basis, and $\bm{k}$-mesh are detailed in our supplementary information (SI).

\subsection{Consistency for Noncollinear LSDA}

\begin{figure}[H]
\centering
\begin{subfigure}[b]{0.45\textwidth}
    \centering
    \includegraphics[width=0.8\textwidth]{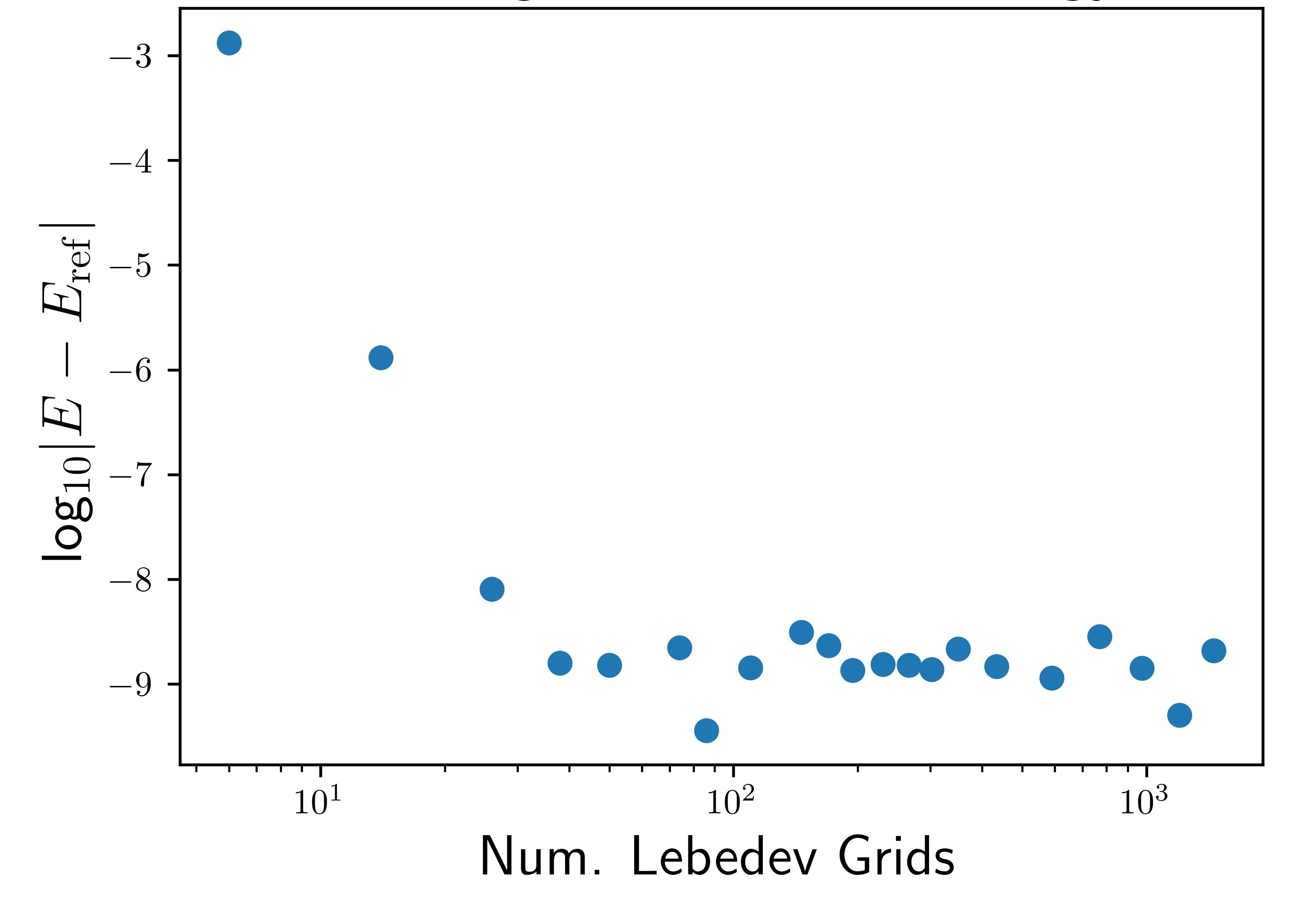}
    \caption{}
    \label{fig:bcc_iron_lda} 
\end{subfigure}
\begin{subfigure}[b]{0.45\textwidth}
    \centering
    \includegraphics[width=0.8\textwidth]{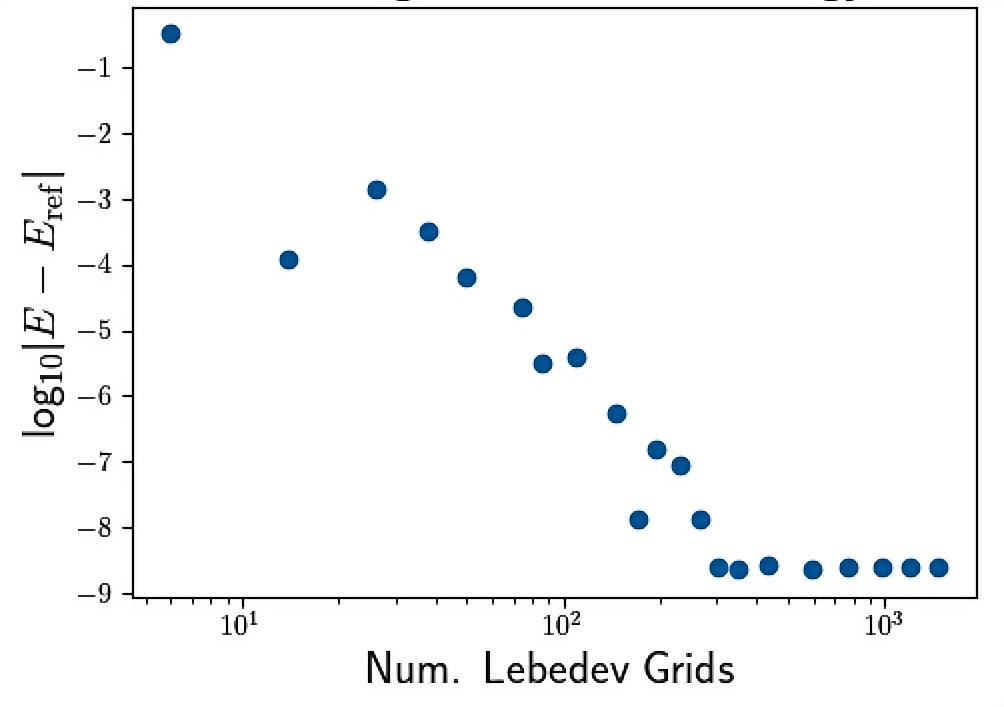}
    \caption{}
    \label{fig:Cr3 LDA}
\end{subfigure}
\caption{(a) Consistency of Noncollinear LSDA on BCC Iron (Energy Unit is eV); (b) Consistency of Noncollinear LSDA on $\mathrm{Cr_3}$ (Energy Unit is eV)}
\end{figure}
Theoretically, the multicollinear approach and locally collinear approach reach the same result for LSDA\cite{PhysRevResearch.5.013036}. We test two approaches on body-centered cubic (bcc) iron and $\mathrm{Cr_{3}}$ cluster (put in a supercell) with Slater exchange \cite{Dirac_1930,Sommerfeld1967ElektronentheorieDM} and PZ correlation 
 \cite{PhysRevB.23.5048}. We present results in Figure \ref{fig:bcc_iron_lda} and Figure \ref{fig:Cr3 LDA} where $E$ is the total energy by the multicollinear approach and $E_{\mathrm{ref}}$ is the total energy by the locally collinear approach. The results indicate that bcc iron is a collinear system and $\mathrm{Cr_{3}}$ is a noncollinear system. We find that the two approaches are consistent with each other for LSDA on the two tested systems. 

The plot for bcc iron shows the convergence of the total energy $E$ calculated using the multicollinear approach relative to the reference energy $E_{\mathrm{ref}}$ obtained from the locally collinear approach. The $x$-axis denotes the quantity of Lebedev grids employed for the generation of unit spin sampling directions in the multicollinear approach. The $y$-axis shows the logarithm of the absolute difference between $E$ and $E_{\mathrm{ref}}$. As the number of Lebedev grids increases, the energy difference $\log_{10}|E-E_{\mathrm{ref}}|$ decreases, indicating better consistency. The curve starts at a higher value for fewer grids (e.g. 10) and gradually decreases, reaching a plateau as the number of grids increases to 1000. This suggests that the multicollinear approach converges to the locally collinear result as the integration becomes more accurate. The small energy differences at higher grid numbers indicate that the multicollinear approach is reliable for such systems. 

Similar to the bcc iron case, the plot for the $\mathrm{Cr_{3}}$ cluster shows the convergence of the total energy $E$ relative to $E_{\mathrm{ref}}$. However, the behavior of the curve is different. The energy difference for the $\mathrm{Cr_{\mathrm{3}}}$ cluster does not decrease as smoothly as in the bcc iron case. The curve exhibits more fluctuations, especially at lower grid numbers (e.g., 10 to 100). This indicates that the multicollinear approach requires a higher number of grids to achieve convergence in noncollinear systems. 

The results underscore the importance of grid sensitivity in the multicollinear approach. For collinear systems like bcc iron, even a moderate number of grids can yield accurate results. However, noncollinear systems like the $\mathrm{Cr_{3}}$ cluster require a significantly higher number of grids to achieve comparable accuracy. Despite the differences in system nature, the global projection and locally collinear approach show consistency within the LSDA framework for both systems. This consistency validates the use of these approaches for studying a range of magnetic systems, provided that the computational parameters are appropriately chosen.

\subsection{Correct Collinear Limit}
\begin{table}[H]
  \centering
  \begin{tabular}{lcccccc}
    \toprule
    \multicolumn{1}{c}{System} & \multicolumn{1}{c}{Type} & \multicolumn{5}{c}{Density Functionals} \\
    \cmidrule(lr){3-7}
    & & PBE  & BLYP & PBEsol & BP86 & OPBE \\
    \midrule
    FCC Nickel & $\Delta E$ & $6.47\times 10^{-11}$ & $1.09\times 10^{-10}$ & $3.01\times 10^{-11}$ & $8.86\times 10^{-10}$ & $3.68\times 10^{-11}$ \\
               & $\Delta M$ & $2.20\times 10^{-9}$ & $1.78\times 10^{-8}$ & $3.00\times 10^{-10}$ & $2.96\times 10^{-6}$ & $3.00\times 10^{-9}$ \\
    \addlinespace
    BCC Iron & $\Delta E$& $2.06\times 10^{-11}$ & $1.63\times 10^{-11}$ & $6.42\times 10^{-11}$ & $1.18\times 10^{-11}$ & $5.37\times 10^{-11}$ \\
             & $\Delta M$ & $1.93\times 10^{-8}$ & $2.30\times 10^{-9}$ & $1.84\times 10^{-8}$ & $1.07\times 10^{-5}$ & $9.10\times 10^{-9}$ \\
    \bottomrule
  \end{tabular}
  \caption{Absolute Total Energy and Atomic Magnetic Moment Deviation between the GKS and UKS (Energy Unit: Rydberg; Magnetic Moment Unit: Bohr Magneton)}
  \label{tab:collinear_limit}
\end{table}

Due to the inherent design of the multicollinear approach, GKS with NCXC is theoretically capable of yielding results equivalent to those obtained from the spin-unrestricted Kohn-Sham (UKS) method with a collinear functional ensemble applied to a collinear system. We conduct the UKS calculation using the collinear functional PBE \cite{PBE}, BLYP \cite{PhysRevB.37.785,PhysRevA.38.3098}, PBEsol \cite{PhysRevLett.100.136406}, BP86 \cite{PhysRevA.38.3098,PhysRevB.33.8822}, and OPBE \cite{HANDY10032001,PBE} provided by Libxc \cite{LEHTOLA20181}, and perform the GKS calculation utilizing the corresponding noncollinear functionals from NCXC. The body-centered cubic (bcc) iron and face-centered cubic (fcc) nickel are selected as the test systems. The absolute energy and atomic magnetic moment deviations between the GKS and UKS are provided in Table \ref{tab:collinear_limit}. The results pertaining to the total energy either meet or exceed the collinear limits as presented by $10^{-10}$ Ry. Likewise, the outcomes concerning the atomic magnetic moment either meet or exceed the collinear thresholds established by $10^{-5}\,\mu_{\mathrm{B}}$.

\subsection{Well-Defined Limit for XC Torque and XC Magnetic Field}

\begin{figure}[H]
    \centering
    \includegraphics[width=0.9\linewidth]{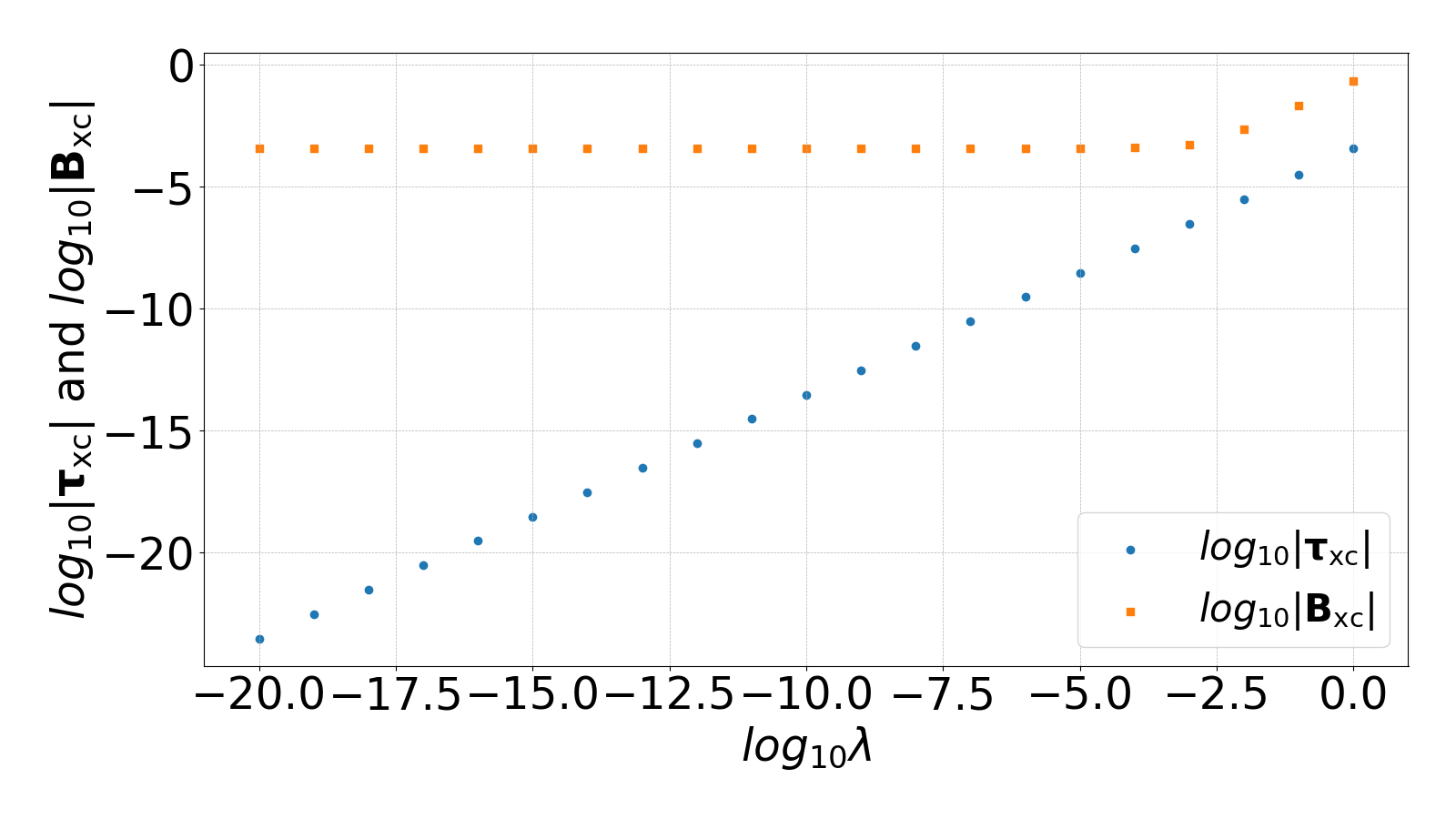}
    \caption{XC Torque and XC Magnetic Field when $\bm{m}$ Approaching Zero}
    \label{fig:torque_b_limit}
\end{figure}

We conduct a numerical test on NCXC, where we use auxiliary input: $n=2.0$, $m_{x}=\lambda$, $m_{y}=0.1\lambda$, $m_{z}=0.01 \lambda$, and others are 0.01. To find out the limit of xc torque and xc magnetic field, $\lambda$ is gradually changed from 1.0 to $10^{-20}$. We set the functional to be the Becke 97-3c by Grimme \emph{et al.} \cite{10.1063/1.5012601}. Figure \ref{fig:torque_b_limit} presents the change of the modulus of the xc torque and xc magnetic field when the spin magnetization vector approaches zero. When $\lambda$ is larger than 0.001, the modulus of the xc magnetic field decreases as $\lambda$ becomes smaller. But the modulus of the xc magnetic field reaches convergence when $\lambda$ is smaller than 0.001, which can be explained by the fact that we don't change the input gradients of the spin magnetization vector and the value of the GGA potential is partially from those gradients. Things are different when it comes to the modulus of the xc torque, where the modulus keeps decreasing as the spin magnetization vector decreases. It is because in this particular setting, the modulus of the xc magnetic field has a non-zero limit while $|\bm{m}|$ has the limit of zero, whose multiplication has the limit of zero.

\section{Application}
\subsection{Noncollinear Magnetism in Spin Spiral}
The high-temperature face-centered cubic (fcc) phase of iron, $\gamma$-Fe, has garnered significant scholarly attention due to its complex and elusive magnetic properties. There exist practical ramifications as well, given that numerous essential iron alloys, including the Invar alloys and several superior-grade stainless steels, exhibit an fcc structure. In 1989, Tsunoda \cite{Tsunoda_1989} successfully stabilized precipitates of $\gamma$-Fe within an fcc copper matrix. This achievement enabled the investigation of the magnetic properties of three-dimensionally confined clusters of $\gamma$-Fe, each with a spherical shape and an average diameter of 50 nanometers. Notably, the ground state was identified as a helical spin density wave, also referred to as a spin spiral, characterized by a specific wave vector. The experiment gives,
\begin{equation}
    \bm{q}_{\mathrm{exp}} = (0.10,0,1)\frac{2\pi}{a}
\end{equation}
In this context, the parameter `$a$' denotes the lattice constant of the conventional face-centered cubic (fcc) copper cell, quantified as 6.822 atomic units (a.u.), which is adopted by the iron precipitates. This wave vector closely approximates a type-I antiferromagnetic configuration, as represented by, 
\begin{equation}
    \bm{q}_{\mathrm{AF}} = (0,0,1)\frac{2\pi}{a}
\end{equation}
In the year 2000, Knöpfle and colleagues \cite{PhysRevB.62.5564} introduced a modified augmented spherical wave (ASW) approach designed to address certain full potential effects that extend beyond the limitations of the atomic sphere approximation (ASA). For lattice constants inferior to 6.75 a.u., this refined methodology produces a novel ground-state $\bm{q}$ vector: \begin{equation}
    \bm{q}_1 = (0.15,0,1)\frac{2\pi}{a}
\end{equation}
This result shows substantial agreement with experimental data, albeit at marginally smaller lattice constants. The outcomes are not significantly modified by the incorporation of the GGA, nor by the deactivation of the intra-atomic noncollinear magnetism applied in this technique. Consequently, the stabilization of this novel ordering vector is attributed to the full potential corrections to the atomic sphere approximation. Nevertheless, when the parameter `$a$' corresponds to the experimentally determined lattice constant of 6.822 a.u., the modified augmented spherical wave (ASW) method stabilizes an incorrect $\bm{q}$ vector: 
\begin{equation}
    \bm{q}_2 =  (0.50,0,1)\frac{2\pi}{a}
\end{equation}

\begin{figure}[H]
    \centering
    \includegraphics[width=0.6\linewidth]{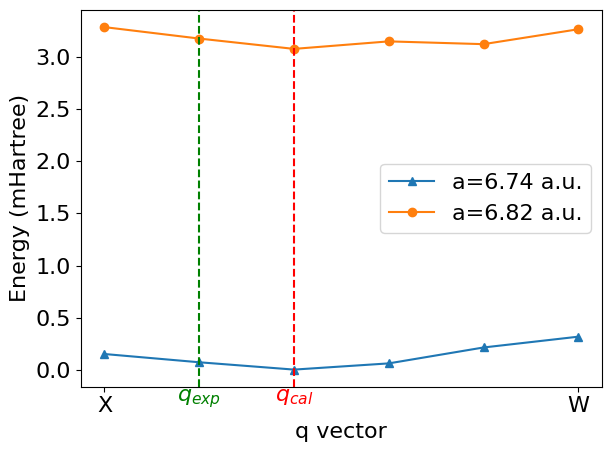}
    \caption{Total Energy of Spin Spiral State Along XW}
    \label{fig:ss}
\end{figure}

In the present study, we employ the full-potential linear augmented plane-wave (FLAPW) method \cite{PhysRevB.24.864} to examine the spin spiral state of $\gamma$-Fe \cite{PhysRevB.69.024415,HEIDE20092678} along XW. X is $(0,0,1)\frac{2\pi}{a}$, and W is $(0,0.5,1)\frac{2\pi}{a}$. Additionally, we utilize the noncollinear LSDA functional PZ \cite{PhysRevB.23.5048}, which is derived from the locally collinear approach. The total energy of spin spiral $\gamma$-Fe is presented in Figure \ref{fig:ss}. For $a = 6.74 \ \mathrm{a.u.}$ and $a = 6.82 \ \mathrm{a.u.}$, we both stabilize an almost quantitatively correct $\bm{q}$ vector:
\begin{equation}
    \bm{q}_{\mathrm{cal}} = (0.20,0,1)\frac{2\pi}{a}
\end{equation}
Elisabeth Sjöstedt and Lars Nordström \cite{PhysRevB.66.014447} previously used the FLAPW method to compute the system using the noncollinear GGA based on the locally collinear approach, potentially with some undisclosed technical modifications to address the singularities. Their results yielded $\bm{q}_{\mathrm{cal}}$ for $a = 6.74 \ \mathrm{a.u.}$, and $\bm{q}_2$ for $a = 6.82 \ \mathrm{a.u.}$. The incorrect determination of the $\bm{q}$ vector when $a = 6.82 \ \mathrm{a.u.}$ is likely attributable to the singularities associated with the noncollinear GGA from the locally collinear approach.

\subsection{Band Structure in Topological Insulators}

Topological insulators (TIs) represent a novel quantum state of matter characterized by an insulating bulk and symmetry-protected metallic surface states, which arise from strong spin-orbit coupling (SOC) and time-reversal symmetry (TRS) \cite{PhysRevLett.95.146802}. The concept of TIs was first theoretically proposed in two-dimensional (2D) systems, such as graphene and HgTe quantum wells, where the quantum spin Hall effect was demonstrated through band inversions driven by SOC \cite{doi:10.1126/science.1133734,doi:10.1126/science.1148047,PhysRevLett.98.106803}. In three-dimensional (3D) TIs, the surface states form odd numbers of massless Dirac cones, with their topological nature determined by the $\mathrm{Z}_2$ invariant calculated from parity eigenvalues at time-reversal-invariant momentum (TRIM) points \cite{9fce9c8ad5154bd3b5e05cda89737fde,PhysRevB.75.121306,PhysRevB.76.045302}. Early experimental efforts focused on Bi$_{1-x}$Sb$_x$ alloys, but their complex surface states and small bulk gaps ($\sim$meV) limited practical applications \cite{9fce9c8ad5154bd3b5e05cda89737fde}. Subsequent theoretical predictions identified stoichiometric crystals like Bi$_2$Se$_3$, Bi$_2$Te$_3$, and Sb$_2$Te$_3$ as robust 3D TIs with simplified surface states and larger bulk gaps \cite{Xia:2009zza,osti_972666,Zhang_2010}. Among these, Bi$_2$Se$_3$ stands out due to its stoichiometric stability, straightforward synthesis, and a substantial bulk bandgap of $\sim$0.3 eV, exceeding the thermal energy scale at room temperature \cite{osti_972666,PhysRevB.74.205113}.

The electronic structure of Bi$_2$Se$_3$ features a single Dirac cone at the \(\Gamma\)-point on the (111) surface, protected by TRS and inversion symmetry \cite{osti_972666}. First-principles calculations reveal that the band inversion between Bi-$p$ and Se-$p$ orbitals at the $\Gamma$-point, driven by strong SOC, underpins its topological nature \cite{Zhang_2010,osti_972666}. The parity analysis of occupied bands at TRIM points confirms a nontrivial $\mathrm{Z}_2$ invariant ($\mathrm{Z}_2 = 1$) \cite{osti_972666}, further validated by surface Green's function calculations showing gapless Dirac states \cite{Zhang_2010}. Experimental verification via angle-resolved photoemission spectroscopy (ARPES) confirmed the existence of these surface states, exhibiting linear dispersion and spin-momentum locking \cite{doi:10.1126/science.1173034}. These properties make Bi$_2$Se$_3$ an ideal platform for exploring topological phenomena and spintronic applications \cite{doi:10.1126/science.1167747}.

For computational modeling, Bi$_2$Se$_3$'s effective Hamiltonian derived from projected atomic Wannier functions (PAWFs) provides a reliable framework to study bulk and surface properties \cite{Zhang_2010}. This approach preserves crystal symmetries and accurately reproduces \emph{ab initio} band structures, enabling efficient simulations of semi-infinite systems or slabs \cite{PhysRevB.56.12847}. The Hamiltonian incorporates SOC as a local atomic term, capturing the critical band inversion mechanism \cite{osti_972666}. Additionally, the penetration depth of surface states ($\sim$2–3 quintuple layers) and spin-resolved Fermi surfaces predicted by this model align with experimental observations, ensuring its utility in predicting emergent quantum effects \cite{doi:10.1126/science.1173034}. These features position Bi$_2$Se$_3$ as a cornerstone material for advancing TI-based device simulations and theoretical studies.

\begin{figure}[H]
\centering
\begin{subfigure}[b]{0.45\textwidth}
    \centering
    \includegraphics[width=0.8\textwidth]{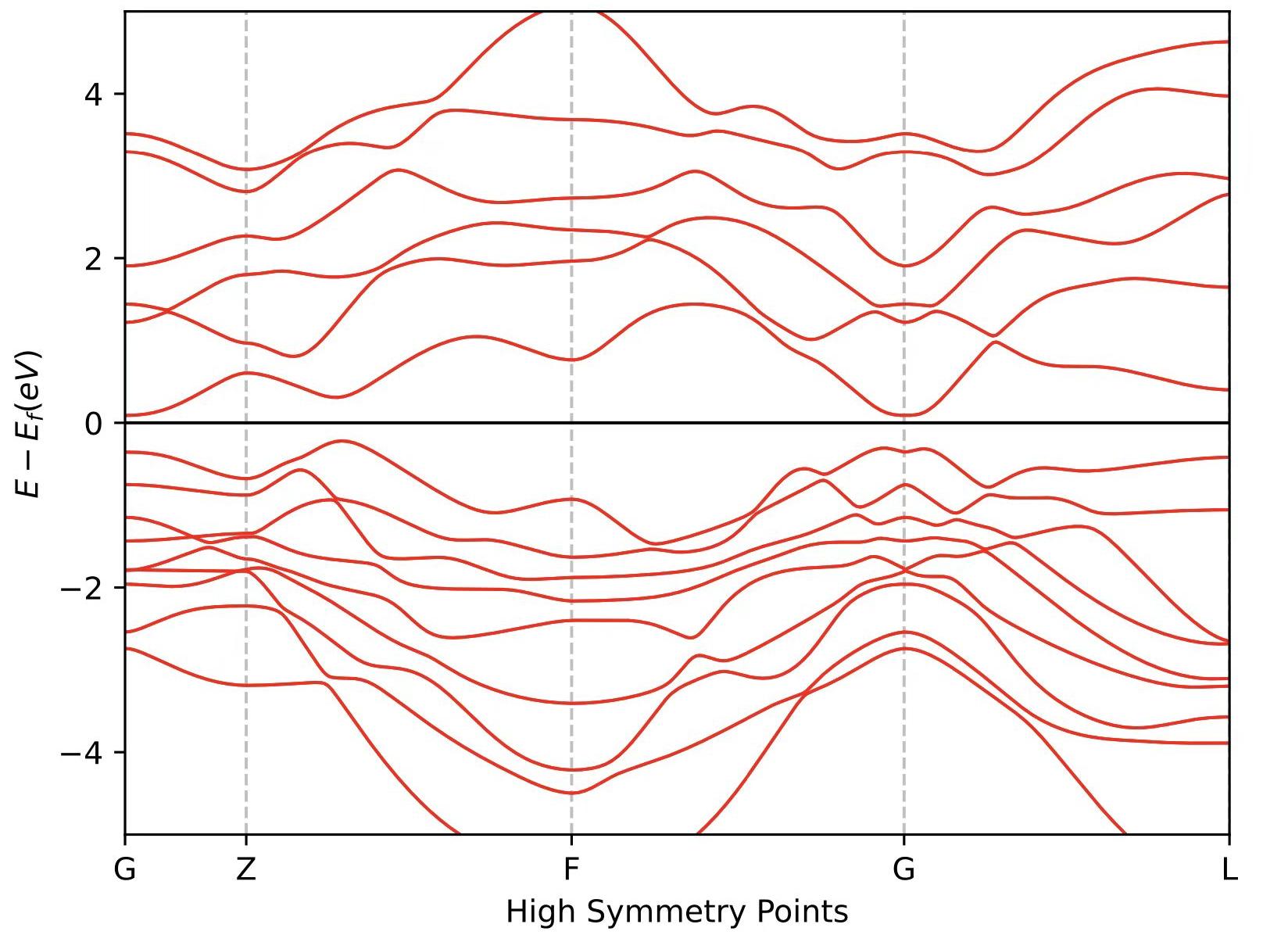}
    \caption{}
    \label{fig:band_stru_soc} 
\end{subfigure}
\begin{subfigure}[b]{0.45\textwidth}
    \centering
    \includegraphics[width=0.8\textwidth]{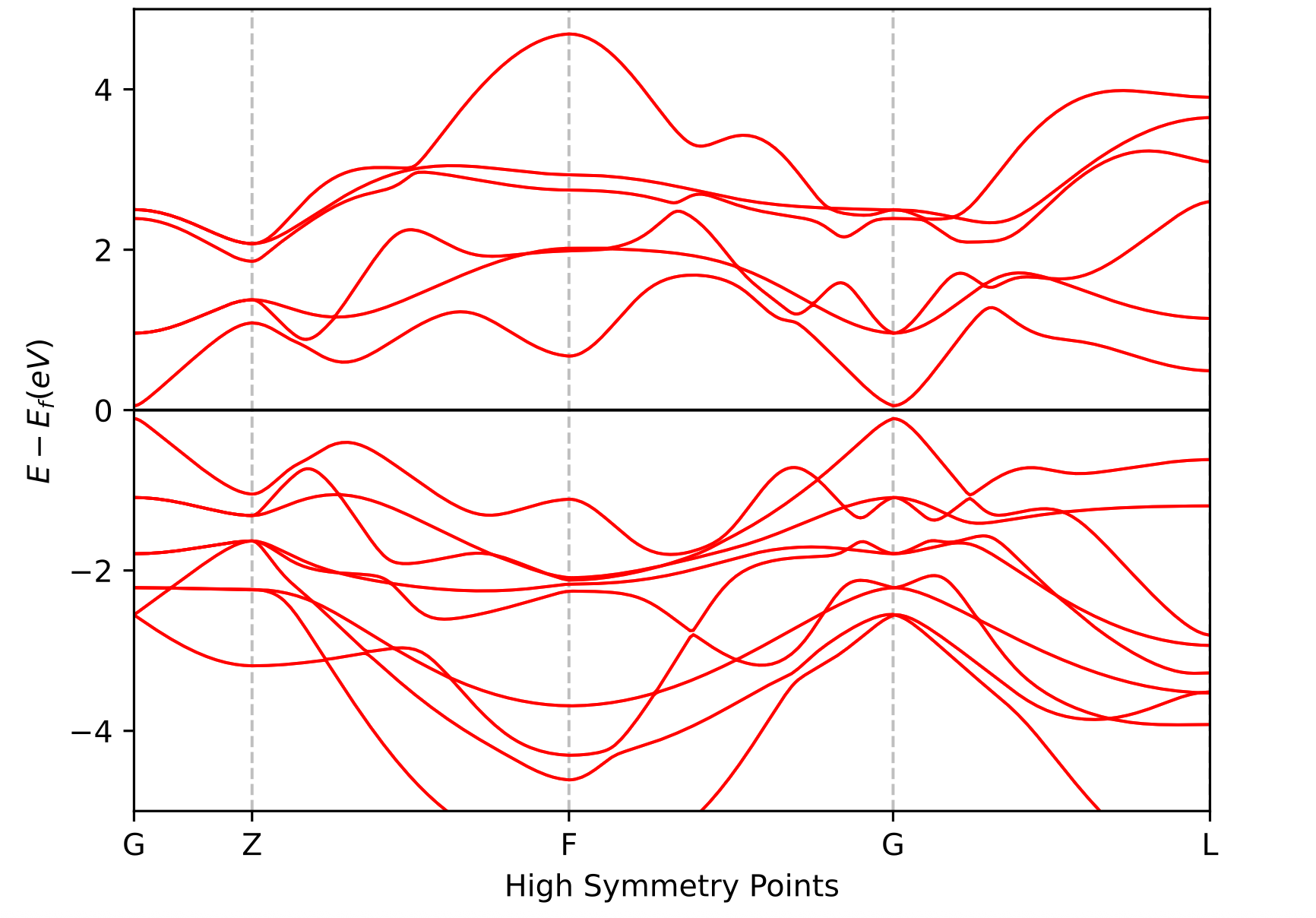}
    \caption{}
    \label{fig:band_stru_nsoc}
\end{subfigure}
\caption{(a) Band Structure of Bi$_2$Se$_3$ Along High Symmetry Points with SOC; (b) Band Structure of Bi$_2$Se$_3$ Along High Symmetry Points without SOC}
\end{figure}

To demonstrate the efficacy of NCXC in the context of topological insulators, we have computed the band structure and band gap of Bi$_2$Se$_3$. Utilizing the PBE functional, the band structures of Bi$_2$Se$_3$ with and without SOC are both demonstrated in Figure \ref{fig:band_stru_soc} and Figure \ref{fig:band_stru_nsoc}. A comparative analysis of the two depicted scenarios reveals that the sole qualitative alteration instigated by the activation of SOC is the emergence of an anti-crossing characteristic near the $\Gamma$ point. This phenomenon signifies an inversion between the conduction and valence bands attributable to SOC effects, thereby implying that Bi$_2$Se$_3$ is a topological insulator.

With the inclusion of SOC, we evaluate the band gap using six distinct functionals: (1) PBE \cite{PhysRevLett.77.3865}; (2) BLYP \cite{PhysRevB.37.785,PhysRevA.38.3098}; (3) BP86; (4) OPBE; (5) B3LYP \cite{10.1063/1.464913} (20\% Hartree-Fock exchange, 8\% Slater \cite{10.1063/1.3129035}, 72\% Becke \cite{PhysRevA.38.3098} for exchange, and 19\% VWN \cite{doi:10.1139/p80-159}, 81\% LYP \cite{PhysRevA.38.3098} for correlation); (6) BHLYP \cite{10.1063/1.464304} (50\% Hartree-Fock exchange and 50\% Slater for exchange, and 100 \% LYP for correlation). The computed band gaps are 0.3201 eV (PBE), 0.3207 eV (BLYP), 0.3196 eV (BP86), 0.3224 eV (OPBE), 0.4905 eV (B3LYP), and 0.6365 eV (BHLYP). It is observed that there is a slight variation in the calculated band gap between the four GGA functionals. With the inclusion of Hartree-Fock exchange, the band gap is progressively increased: B3LYP, with 20\% Hartree-Fock exchange, yields a band gap of 0.4905 eV, whereas BHLYP, with 50\% Hartree-Fock exchange, results in a band gap of 0.6365 eV. 

\subsection{Band Gap in Semiconducting Inorganic Materials}
The band gap ($E_{\mathrm{g}}$) represents an essential parameter that is intrinsically linked to the applicability of materials in the realms of optical, electronic, and energy applications. For example, in photovoltaic systems, materials featuring a direct $E_{\mathrm{g}}$ of approximately 1.3 eV \cite{10.1063/1.1736034,doi:10.1021/acs.chemmater.9b00708}, in alignment with the Shockley-Queisser limit, are preferred as they serve as optimal photo-absorbers, thereby enhancing solar cell efficiency. In the domain of power electronics, semiconductors possessing a $E_{\mathrm{g}}$ of 3 eV or greater are utilized to withstand high electric fields \cite{gorai2019computational}. To augment the figure of merit in thermoelectric systems, the selection of materials with a $E_{\mathrm{g}}$ of 10 $k_{\mathrm{B}}T_{\mathrm{op}}$, where $k_{\mathrm{B}}$ and $T_{\mathrm{op}}$ denote the Boltzmann constant and operating temperature, respectively, is advisable \cite{PhysRevB.49.4565}. Considering the pivotal role of $E_{\mathrm{g}}$, the establishment of a comprehensive database of $E_{\mathrm{g}}$ across an extensive range of materials can significantly streamline the material selection process for specific applications by efficiently eliminating less suitable candidates. At present, widely recognized material databases such as the Materials Project (MP) \cite{10.1063/1.4812323}, the Automatic Flow of Materials Discovery Library (AFLOW) \cite{CURTAROLO2012227}, and the Open Quantum Materials Database (OQMD) \cite{saal2013materials} offer theoretical $E_{\mathrm{g}}$ for a collection encompassing up to one million inorganic compounds.

\begin{table}[H]
\centering
\begin{tabular}{llcccccccc}
\hline
\textbf{Name}       & \textbf{ICSD} & \textbf{Exp.} & \textbf{PBE} & \textbf{BLYP} & \textbf{BP86} & \textbf{OPBE} & \textbf{MP} & \textbf{OQMD} & \textbf{AFLOW} \\
\hline
IrSb$_3$          & 640958 & 1.18 & 0.45 & 0.48 & 0.46 & 0.22 & 0.05 & 0.21 & 0.00 \\
Cu$_3$AsSe$_4$    & 610361 & 0.88 & 0.01 & 0.00 & 0.01 & 0.02 & 0.00 & 0.00 & 0.00 \\
RhAs$_3$          & 34052  & 0.85 & 0.25 & 0.20 & 0.23 & 0.26 & 0.00 & 0.00 & 0.00 \\
RhSb$_3$          & 650248 & 0.80 & 0.27 & 0.22 & 0.25 & 0.31 & 0.00 & 0.15 & 0.00 \\
AgAlTe$_2$        & 28746  & 2.35 & 0.99 & 1.00 & 0.97 & 1.24 & 1.05 & 1.20 & 1.36 \\
\hline
\multicolumn{3}{c}{MD/eV}   & -0.82 & -0.83 & -0.83 & -0.80 & -0.99 & -0.90 & -0.94 \\
\multicolumn{3}{c}{MAD/eV}  & 0.82  & 0.83  & 0.83  & 0.80  & 0.99  & 0.90  & 0.94  \\
\multicolumn{3}{c}{RMSE/eV} & 0.87  & 0.88  & 0.88  & 0.83  & 1.01  & 0.91  & 0.95  \\
\hline
\end{tabular}
\caption{Calculated and Experimental Band Gaps with Multiple Functionals}
\label{tab:band_gap}
\end{table}
In general, the present database provides results $E_{\mathrm{g}}$ that are in close agreement with experimental measurements. However, certain materials display significant discrepancies of $\geq$ 0.5 eV. We arbitrarily selected five of these materials. The experimental data for IrSb$_3$, Cu$_3$AsSe$_4$, RhAs$_3$, and RhSb$_3$ were retrieved from a CRC Handbook \cite{lide2005crc}, while the data for AgAlTe$_2$ \cite{tell1974some} was derived from the respective research articles. The five materials were analyzed using PBE, BLYP, BP86, OPBE, with their calculated and experimental band gaps presented in Table \ref{tab:band_gap}. The tabulated data include the calculated band gaps incorporating spin-orbit coupling (SOC).

The following analysis is based on band gaps with SOC. Cu$_3$AsSe$_4$ represents a deviation from our predictive method, with an experimental band gap value measured at 0.88 eV, while theoretical predictions yield values approaching zero. The accurate theoretical comprehension of the electronic properties of these materials is impeded by the involvement of Cu d electrons. Calculations based on density functional theory utilizing the local density approximation or generalized gradient approximation frequently yield qualitatively inaccurate electronic properties for these materials, particularly in similar narrow-gap systems. An examination of the band gaps for approximately 20 Cu-based semiconductors, computed using the $\mathrm{mBJ}+U$ method, indicates that the findings concur with reliable values within a margin of $\pm$ 0.2 eV. \cite{10.1063/1.4828864} 
For the left four materials, our method can yield qualitatively accurate results, whereas predictions derived from the three datasets may inaccurately result in a zero value. From a statistical perspective, our methodology demonstrates superior accuracy in terms of mean deviation, mean absolute deviation, and root mean square error. In this context, the four GGA functionals exhibit almost equal performance.

\section{Conclusion}
The multicollinear approach is theoretically applicable to any category of density functionals. Similarly, our operator formalism, in principle, can be applied to any form of density functional. It means that our theory can be used beyond LSDA, GGA and meta-GGA. In this study, we have explicitly implemented the formulations for noncollinear LSDA and GGA. Implementation of noncollinear orbital-dependent mGGA is anticipated via the noncollinear optimal effective potential method, which will be investigated in our future research. Beyond the ground-state GKS calculation, our novel formalism facilitates the independent implementation of noncollinear kernels for spin-flip TDDFT without the necessity of considering basis functions, thereby directly endorsing the Sternheimer formalism \cite{PhysRev.84.244,10.1063/1.2733666}. Furthermore, our formalism directly provides the exchange and correlation torque, thereby offering enhanced understanding in real-time TDDFT and the Landau-Lifshitz-Gilbert equation concerning spin dynamics.

\begin{acknowledgement}
Xiaoyu Zhang thanks valuable discussions with Yunlong Xiao and Yixiao Chen. We thank the technical suggestions and help from Mohan Chen, Daye Zheng, Zuxin Jin, Hao Li and Zhichen Pu.
\end{acknowledgement}

\section*{\textbf{Author Information}}

\textbf{Corresponding Author}
\begin{itemize}[leftmargin=16pt, nosep]
    \item Xiaoyu Zhang - College of Chemistry and Molecular Engineering, Peking University, Beijing, 100871, P. R. China.
    
    \url{https://orcid.org/0009-0009-4178-3519}; 
    Email: \href{mailto:zhangxiaoyu@stu.pku.edu.cn}{\texttt{zhangxiaoyu@stu.pku.edu.cn}}
\end{itemize}

\textbf{Authors}
\begin{itemize}[leftmargin=16pt, nosep]
    \item Xiaoyu Zhang - College of Chemistry and Molecular Engineering, Peking University, Beijing, 100871, P. R. China;
    \item Taoni Bao - School of Physics, Peking University, Beijing, 100871, P. R. China; Center for Applied Physics and Technology, Peking University, Beijing, 100871, P. R. China.
\end{itemize}

\textbf{Author Contributions}\\
Xiaoyu Zhang and Taoni Bao are joint first authors for this work.

\textbf{Notes}
\begin{itemize}[leftmargin=16pt, nosep]
    \item The authors declare no competing financial interest.
\end{itemize}

\bibliography{main}

\section*{Appendix A: Definition of Discrete Matrix-Valued Multiplicative Operator}
Let $\Gamma$ be a discrete lattice (e.g., $\Gamma = \mathbb{Z}^d$ for a $d$-dimensional lattice or a finite subset). 
The Hilbert space of two-component wave functions is denoted as $\mathcal{H} = \ell^2(\Gamma, \mathbb{C}^2)$, 
where a wave function $\psi \in \mathcal{H}$ satisfies:
\[
\|\psi\|^2 = \sum_{\bm{r} \in \Gamma} \|\psi(\bm{r})\|^2_{\mathbb{C}^2} = \sum_{\bm{r} \in \Gamma} \left( |\psi_1(\bm{r})|^2 + |\psi_2(\bm{r})|^2 \right) < \infty.
\]

An operator $\hat{O}: \mathcal{H} \to \mathcal{H}$ is called a \textit{discrete matrix-valued multiplication operator} 
if there exists a matrix-valued function $M: \Gamma \to \mathbb{C}^{2 \times 2}$ such that for all $\psi \in \mathcal{H}$ 
and all lattice points $\bm{r} \in \Gamma$, the action of $\hat{O}$ is given by:
\[
(\hat{O} \psi)(\bm{r}) = M(\bm{r})  \psi(\bm{r}) = 
\begin{pmatrix} 
M_{11}(\bm{r}) & M_{12}(\bm{r}) \\ 
M_{21}(\bm{r}) & M_{22}(\bm{r}) 
\end{pmatrix}
\begin{pmatrix} 
\psi_1(\bm{r}) \\ 
\psi_2(\bm{r}) 
\end{pmatrix}.
\]
Equivalently, in component form:
\[
(\hat{O} \psi)(\bm{r}) = 
\begin{pmatrix} 
M_{11}(\bm{r}) \psi_1(\bm{r}) + M_{12}(\bm{r}) \psi_2(\bm{r}) \\ 
M_{21}(\bm{r}) \psi_1(\bm{r}) + M_{22}(\bm{r}) \psi_2(\bm{r}) 
\end{pmatrix}.
\]

The operator has the following key properties:
\begin{enumerate}
    \item \textbf{Locality:} 
    $(\hat{O} \psi)(\bm{r})$ depends only on the value of $\psi$ at $\bm{r}$, 
    with no coupling to other lattice points.
    
    \item \textbf{Non-differential:} 
    The operator contains no differential, gradient, or finite-difference components 
    (e.g., no $\nabla$, $\partial_x$, or discrete Laplacian terms).
    
    \item \textbf{Boundedness:} 
    If $\sup\limits_{\bm{r} \in \Gamma} \|M(\bm{r})\|_{\mathrm{op}} < \infty$, 
    then $\hat{O}$ is bounded with $\|\hat{O}\| \leq \sup \|M(\bm{r})\|_{\mathrm{op}}$.
    
    \item \textbf{Reduction to scalar case:} 
    If $M(\bm{r}) = \begin{pmatrix} a(\bm{r}) & 0 \\ 0 & b(\bm{r}) \end{pmatrix}$, 
    then $\hat{O}$ decomposes into two scalar-multiplicative operators.
\end{enumerate}

\end{document}